# Vibrational states of a water molecule in a nano-cavity of beryl crystal lattice

Elena S. Zhukova[1,2,3,a], Victor I. Torgashev[4], Boris P. Gorshunov[1,2,3], Vladimir V. Lebedev[3,5], Gil'man S. Shakurov[6], Reinhard K. Kremer[7], Efim V. Pestrjakov[8], Victor G. Thomas[9], Dimitry A. Fursenko[9], Anatoly S. Prokhorov[1,3] and Martin Dressel[2]

[1] A.M. Prokhorov General Physics Institute, Russian Academy of Sciences, Vavilov Street 38, 119991 Moscow, Russia

[2] 1. Physikalisches Institut, Universität Stuttgart, Pfaffenwaldring 57, 70550 Stuttgart, Germany

[3] Moscow Institute of Physics and Technology, 141700 Dolgoprudny, Moscow Region, Russia

[4] Faculty of Physics, Southern Federal University, 344090 Rostov-on-Don, Russia

[5] Landau Institute for Theoretical Physics, Russian Academy of Sciences, Chernogolovka, Moscow Region, Russia

[6] Kazan Physical-Technical Institute, Russian Academy of Sciences, 10/7 Sibirsky Trakt, 420029 Kazan, Russia

[7] Max-Planck-Institut für Festkörperforschung, Heisenbergstrasse 1, 70569 Stuttgart, Germany

[8] Institute of Laser Physics, Russian Academy of Sciences, 13/3 Ac. Lavrentyev's Prosp., 630090 Novosibirsk, Russia

[9] Institute of Geology and Mineralogy, Russian Academy of Sciences, 3 Ac. Koptyug's Prosp., 630090 Novosibirsk, Russia

Low-energy excitations of a single water molecule are studied when confined within a nano-size cavity formed by the ionic crystal lattice. Optical spectra are measured of manganese doped beryl single crystal Mn:Be$_3$Al$_2$Si$_6$O$_{18}$, that contains water molecules individually isolated in 0.51 nm diameter voids within the crystal lattice. Two types of orientation are distinguished: water-I molecules have their dipole moments aligned perpendicular to the *c* axis, and dipole moments of water-II molecules are parallel to the *c*-axis. The optical conductivity *σ(v)* and permittivity *ε′(v)* spectra are recorded in terahertz and infrared ranges, at frequencies from several wavenumbers up to $v = 7000$ cm$^{-1}$, at temperatures 5 K – 300 K and for two polarizations, when the electric vector **E** of the radiation is parallel and perpendicular to the *c*-axis. Comparative experiments on as-grown and on dehydrated samples allow to identify the spectra of *σ(v)* and *ε′(v)* caused exclusively by water molecules. In the infrared range, well-known internal modes $v_1$, $v_2$ and $v_3$ of the H$_2$O molecule are observed for both polarizations, indicating the presence of water-I and water-II molecules in the crystal. Spectra recorded below 1000 cm$^{-1}$ reveal a rich set of highly anisotropic features in the low-energy response of H$_2$O molecule in a crystalline nano-cavity. While for **E**∥*c* only two absorption peaks are detected, at ~90 cm$^{-1}$ and ~160 cm$^{-1}$, several absorption bands are discovered for **E**⊥*c*, each consisting of narrower resonances. The bands are assigned to librational (400 cm$^{-1}$ – 500 cm$^{-1}$) and translational (150 cm$^{-1}$ – 200 cm$^{-1}$) vibrations of water-I molecule that is weakly (via hydrogen bonds) coupled to the nano-cavity "walls". A model is presented that explains the "fine structure" of the bands by splitting of the energy levels due to quantum tunneling between the minima in a six-well potential relief felt by a molecule within the cavity.

---

[a] Author to whom correspondence should be addressed. Electronic mail: zhukovaelenka@gmail.com. Telephone: 007 499 503 8212.

## I. INTRODUCTION

One of the frontiers in present solid state sciences deals with phenomena occurring at nano-scale. If the dimension of a material becomes comparable to certain intrinsic length scales (electron or phonon mean free path, the de Broglie wavelength, a correlation length of some kind of collective interaction, etc.), qualitatively new effects can be observed that are not manifested in the material in its bulk form.[1, 2] Nano-size phenomena may also be observed if the amount of matter is drastically reduced – down to the quantity of only a few molecules or atoms. Although interactions between the members of the ensemble are of the same nature as in the bulk, *'nanofication'* can lead to spatial arrangements of the bonds different from that in the bulk and can transform the system into novel states. The diversity of the system properties is further enriched by introducing extended interfaces disrupting bonds locally, or by finite-size effects if a system is confined e.g. into nano-pores, nano-wires, or nano-tubes. Understanding the nature of the emerging phases and their relations to the physical, chemical, geometrical and morphological characteristics of the environment is presently still at its infancy. The systematic investigation of such nano-size effects may open exciting areas in modern fundamental and applied research.

In such studies, particular attention is focused on water molecules since water is crucial for sustaining life and for activity of all biological systems.[3, 4] Although isolated $H_2O$ molecules seem to be rather simple, bulk water is one of the least understood liquids. It displays extraordinary properties that result from labile and highly-directional hydrogen bonds (H-bonds), relatively small moment of inertia of rotation and large dipole moment.[5, 6, 7, 8, 9] The water-confinement effects are so peculiar that *confined* water is sometimes called the "fifth state of water"[10]. Alternatively, *confined* water is labeled as zeolitic water, one-dimensional water, interfacial water, nanotube water (ice nanotube), biological water, quantum water. For example, water molecules located in one-dimensional carbon natotubes exhibits very special collective excitations[11] and new quantum states,[12] together with the formation of firmly connected one-dimensional hydrogen-bonded water "wires"[13] or tube-like structures.[14] In such objects one can observe anomalously soft molecular dynamics[15] or anisotropic and rapid molecular/proton transport.[16, 17, 18, 19, 20] Advanced experimental techniques have been developed to open and close the spherical fullerene molecule, and to encapsulate a water molecule into a highly symmetrical cage of $C_{60}$[21, 22] and also $C_{70}$. In addition, aggregation of more than one $H_2O$ molecule can be trapped within nano-sized cavities of zeolites, clays, silica gels, globular photonic crystals and other natural and synthesized frameworks (see [23, 24] and references therein) providing a fascinating playground for supramolecular physics and chemistry. Nano-scale water plays a vital role in biology where it controls the structure, stability, functionality and reactivity of bio-molecules. Water molecules at the interfaces (hydration layers) of biological complexes, in channels of membranes, in solutions, participate in the hydrophobic/hydrophilic interactions, replication and transcription processes. [25, 26, 27, 28, 29, 30, 31, 32, 33]

The properties revealed by water at the nano-scale are extremely intricate, especially in complex biological systems. In order to get more insight into the fundamental mechanisms, studies on very basic objects are necessary. Here we have chosen a simple model system: crystals of the beryl group with chemical composition given by the formula $Be_3Al_2Si_6O_{18}$. Beryl crystallizes with a honeycomb structure (space group *P6/mcc*) (Fig. 1a) consisting of stacked six-membered rings of $SiO_4$ tetrahedra that leave relatively large open channels oriented along the crystallographic *c* axis.[34, 35] The channels include constrictions ("bottlenecks") of approximately 0.28 nm diameter separating larger cavities of 0.51 nm in diameter. Crystals that are grown in water environment (natural water-containing silicate melts or hydrothermal solutions) contain water that enters the framework of the crystal lattice in such a way that *single* $H_2O$ molecules reside within the cavities, importantly - *in a well-defined symmetrical crystallographic environment*. The $H_2O$ molecules are held in two orientations relative to the *c*-axis (Fig. 1b, c): water-I molecules have the vector connecting two protons directed parallel to the *c*-axis (electric dipole moment perpendicular to *c*), whereas in water-II molecules the H-H vector is perpendicular to the *c*-axis (dipole moment parallel to *c*) (see for instance Ref. [36]). Water-II molecules are found in beryl crystals containing alkali ions (Na, K), which are located at the bottlenecks. It is Coulomb interaction with these ions that turns the water-II molecules 90 degrees relative to the water-I molecules. The water-I molecules are loosely tied to surrounding oxygen atoms by weak H-bonds and are thus more labile in the cavities, they can even be considered as *nearly* free.[36, 37] Yet the weak coupling to the cavities walls leads to *qualitatively new* dynamical characteristics of the molecules, as compared to those of free $H_2O$ molecules.

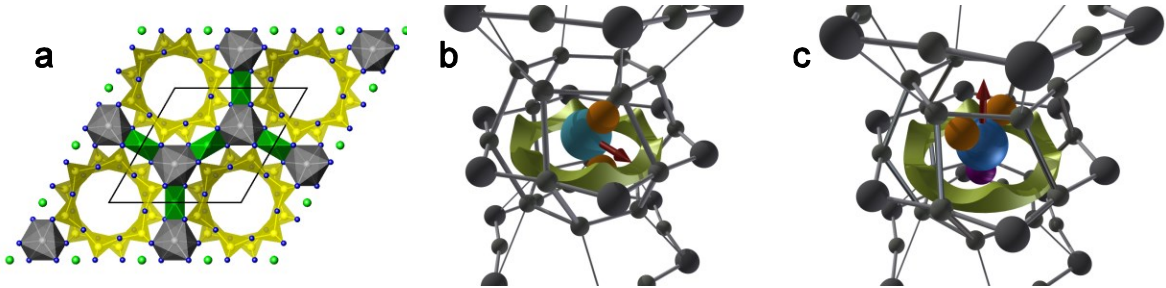

FIG. 1. (Color online). (a) Honeycomb crystal structure (space group *P*6/*mcc*) of beryl in the plane perpendicular to the *c*-axis. Stacked six-membered rings of $SiO_4$ form open nano-channels extending along the c-direction. They contain cavities of 0.51 nm diameter connected by bottlenecks of 0.28 nm that can be clogged by alkali ions (Na or K). Crystal water molecules within the cavities are oriented with the dipole moment either perpendicular (b) or parallel (c) to the *c*-direction. The water-II molecules are rotated by 90° relative to the water-I molecules due to the Coulomb interaction with the positively charged alkali ions. Hydrogen bonds weakly connect the water molecules to surrounding oxygen atoms of the $SiO_4$ cavity, as indicated by the green belt that resembles the periodic potential. The red arrows indicate the dipole moments of the $H_2O$ molecules, which are subject to librations and translations.

Indications of peculiar low-energy vibrational states of confined water-I are seen in the infrared spectra of beryl. Here, the well-known intramolecular modes ($v_1$ = 3656.65 cm$^{-1}$, $v_2$ = 1594.59 cm$^{-1}$, $v_3$ = 3755.79 cm$^{-1}$ for free $H_2O$[38]) are observed [37, 39, 40, 41, 42, 43, 44, 45] and accompanied by satellite peaks. These peaks are attributed to a mixture of $v_1$, $v_2$ and $v_3$ with the so-called external, librational (L) and translational (T) vibrations of $H_2O$ molecule within the beryl nano-cavities.[37, 44] According to the experimental spectra obtained by Kolesov *et al.*[37, 44] one expects a very rich set of such T- and L- type excitations with eigenfrequencies ranging from approximately 1000 cm$^{-1}$ down to only few wavenumbers, i.e. down to excitations energies of order of a meV. In an attempt to *directly* register such lowest-energy external vibrations, Raman measurements have been performed down to frequencies as low as 100 cm$^{-1}$. A series of narrow and broad resonances has been detected.[37] A water-related band was observed around 200 cm$^{-1}$ that was split at 5 K into several satellites. An attempt was made to assign the satellites to combinations of the $v_3$ vibration with Stokes- and anti-Stokes components $v_3 + n\omega$ ($n$ = 1, 2, 3, …) with $\omega$ = 6.3 cm$^{-1}$ [44] or $\omega$ = 9 cm$^{-1}$,[37] or, alternatively, to "particle in the box"-like motions of the $H_2O$ molecule within the beryl nano-cavity.[45] However, severe differences between this assignment and the experiment remain. It is worth noting that similar low-energy excitations were observed in infrared and Raman spectra of other systems that contain nano-confined water, e.g. hemimorphite ($Zn_4Si_2O_7(OH)_2 \cdot H_2O$),[46] natrolite $Na_{16}[Al_{16}Si_{24}O_{80}] \cdot 16H_2O$ or scolecite $Ca_8[Al_{16}Si_{24}O_{80}] \cdot 24H_2O$.[46] In some cases, low-frequency measurements allowed direct observation of excitations at frequencies down to 50 cm$^{-1}$ - 100 cm$^{-1}$ and to assign them to translational external vibrations of $H_2O$ molecule or to combination of such translations with lattice phonons.

Above described results clearly demonstrate the rich nature of low-energy vibrational states of single water molecule caused by its confinement inside nano-sized cavities of beryl and similar compounds. However, no systematic study of the corresponding low-energy dynamics of $H_2O$ molecules has been performed yet due to difficulties connected with a conclusive assignment of the optical absorption spectra and the problem to distinguish water and crystal lattice phonon modes. Additionally, the special instrumental requirements when extending into the very low-frequency region of the spectra aggravated such experiments. In this work we aimed at first detailed and systematic investigation of the vibrational states of a single water molecule encapsulated in the nano-cavity of the beryl crystal lattice.[47] We employed optical spectroscopy techniques in an extremely wide frequency range from $v$ = 7000 cm$^{-1}$ down to few wavenumbers (2 cm$^{-1}$) in the temperature range from room temperature down to 5 K. We found that already a weak H-bond linking of $H_2O$ molecule to the cavity walls results in a rich set of highly anisotropic soft energy states connected with the movements of the molecule as a whole within the cavity. We present a model that accounts for the observed set of energy levels as coming from translational and librational molecular oscillations enriched by quantum tunneling between the states in a six-well potential relief felt by a molecule within the crystalline cavity.

## II. EXPERIMENT

Beryl single crystals were grown in stainless steel autoclaves according to the standard hydrothermal growth method[48] at a temperature of 600°C and under pressure of 1.5 kbar by a recrystallization of natural beryl to a seed crystal. The chemical composition of the crystals amounts to the mass %: $SiO_2$ – 65.79, $Al_2O_3$ – 17.32, BeO – 13.75, $Fe_2O_3$ – 1.30, MnO – 0.09, $Li_2O$ – 0.15, $H_2O$ – 1.93, with some traces of $Na_2O$, $K_2O$, CuO and MgO. Recalculation of these values to the crystallographic formula gives:[49] $(Be_{2.988}Li_{0.012})$ $(Al_{1.865}Fe^{3+}_{0.090}Mn^{3+}_{0.007}Si_{0.038})$ $(Si_{5.971}Be_{0.029})((H_2O)_{0.532}Li_{0.043})$.

For optical measurements a beryl crystal of about one cubic centimeter size was oriented by X-rays and cut in thin slices with the crystallographic *c*-axis within their planes. This geometry allows us to measure the optical response in two principal polarizations with the electrical field vector ***E*** of the probing radiation oriented parallel and perpendicular to the *c*-axis. Throughout the text these orientations are referred to as parallel (***E***||*c*) and perpendicular (***E***⊥*c*), respectively.

Optical measurements were performed using two kinds of spectrometers: in the infrared range, a standard Fourier transform spectrometer (Bruker IFS-113V) was used to measure the spectra of reflection $R(v)$ and transmission $Tr(v)$ coefficients at frequencies from 20 - 30 $cm^{-1}$ up to 7000 $cm^{-1}$. The reflectivity spectra were recorded on samples of about 1 mm thickness. For the transmissivity measurements thinner (about 100 micrometers) samples were prepared. The same specimens were measured at lower frequencies (down to few cm$^{-1}$) with the help of a quasioptical spectrometer based on monochromatic and continuously frequency tunable radiation generators – backward-wave oscillators (BWOs). The BWO-spectrometer is described in details in.[50, 51] Employing a Mach-Zehnder interferometric arrangement allows us to determine directly (without performing a Kramers-Kronig transformation) the complex optical conductivity $\sigma^*=\sigma_1+i\sigma_2$ (or dielectric permittivity $\varepsilon^*=\varepsilon'+i\varepsilon''$) at frequencies from $v = 1$ cm$^{-1}$ up to $v = 50$ cm$^{-1}$, in the temperature interval from 5 K to 300 K.

The optical experiments were complemented by measurements of the heat capacity in the relaxation method employing a PPMS system (Quantum Design). The heat capacities of a minute amount of Apiezon N vacuum grease and the sample platform were determined in a preceding run and subtracted from the total heat capacities.

## III. RESULTS

Fig. 2a, b exhibits typical broadband spectra of transmission $Tr(v)$ and the reflection $R(v)$ coefficients of a beryl crystal measured with polarization $\boldsymbol{E}\perp c$ at liquid helium temperature. The spectra contain a rich sets of absorption lines identified as minima in $Tr(v)$ and as characteristic dispersions in $R(v)$. The lowest-frequency resonance absorption is observed between 10 cm$^{-1}$ and 30 cm$^{-1}$, which is re-plotted on a linear scale in Fig. 2c (note the logarithmic ordinate) where the spectra of the sample with and without (see below) water molecules are compared. For both polarizations no further resonances were detected down to 2 cm$^{-1}$. The narrow absorption lines at frequencies above 1000 cm$^{-1}$ correspond to the well-known intramolecular modes $v_1$, $v_2$, $v_3$ of the water molecules. They are indicated by arrows in Fig. 1a. In accordance with previous measurements [37, 44] these modes are accompanied by satellites (shown in more details below) due to coupling to lower-energy external H$_2$O vibrations. The structures observed between 100 cm$^{-1}$

and 1000 cm$^{-1}$ are composed of the mixed response of the crystal lattice (phonons) and of water molecules in the cavities of the crystal.

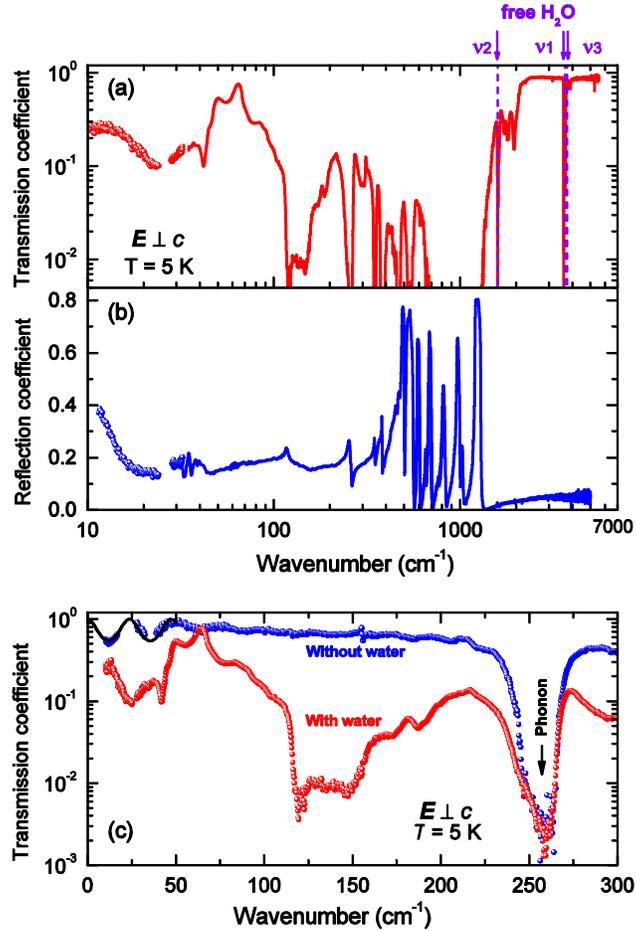

FIG. 2. (Color online). Spectra of transmission (a, sample thickness 102 μm) and reflection (b, sample thickness 1.34 mm) coefficients of beryl measured for polarization $\boldsymbol{E}\perp c$ at T = 5 K. Arrows and dashed lines at the upper panel indicate frequencies of intramolecular stretching and bending modes ($v_1$, $v_2$, $v_3$) of the free water molecule. Dots in panel (a) show the spectra collected with the terahertz BWO-spectrometer and the solid lines in panels (a) and (b) by infrared Fourier transform spectrometer. Dots in panel (b) correspond to reflectivity values calculated basing on directly measured terahertz spectra of real $\varepsilon'$ and imaginary $\varepsilon''$ parts of dielectric permittivity. Panel (c) shows spectra of the transmission coefficients of beryl (sample thickness 102 μm) measured for polarization $\boldsymbol{E}\perp c$ at T = 5 K before and after dehydration (annealing in vacuum at 1000°C for 24 hours). The solid line at low frequencies shows an example of model fitting of the spectra (see text). Arrow shows absorption due to phonon.

In order to extract characteristic parameters of each resonance and to trace their temperature evolution, we have analyzed the measured $Tr(v)$ and $R(v)$ spectra using the well-known Fresnel expressions for complex transmission $Tr^*$ and reflection $R^*$ coefficients of a plane-parallel layer [52, 53]

$$Tr^* = \frac{T_{12}T_{21}\exp(i\delta)}{1+T_{12}T_{21}\exp(2i\delta)}, \tag{1}$$

$$R^* = \frac{R_{12}+R_{21}\exp(2i\delta)}{1+R_{12}R_{21}\exp(2i\delta)}, \tag{2}$$

where

$$T_{pq} = t_{pq}\exp(i\varphi^{T}_{pq}),\ R_{pq} = r_{pq}\exp(i\varphi^{R}_{pq}),\ t^2_{pq} = \frac{4(n^2_p+k^2_p)}{(k_p+k_q)^2+(n_p+n_q)^2},$$

$$r^2_{pq} = \frac{(n_p-n_q)^2+(k_p-k_q)^2}{(k_p+k_q)^2+(n_p+n_q)^2},\ \varphi^{T}_{pq} = arctg\left(\frac{k_p n_q - k_q n_p}{n^2_p+k^2_p+n_p n_q+k_p k_q}\right),$$

and

$$\varphi^{R}_{pq} = arctg\left[\frac{2(k_p n_q - k_q n_p)}{n^2_p+k^2_p-n^2_q-k^2_q}\right]$$

are the Fresnel coefficients for the interfaces "air-to-sample". The indixes $p, q = 1, 2$ correspond to the air (refractive index $n_1 = 1$, extinction coefficient $k_1 = 0$) and to the material of the sample ($n_2, k_2$), $\delta = 2\pi d/\lambda(n_2 + ik_2)$, respectively. $d$ is the thickness of the sample, $\lambda$ is the wavelength of the radiation. Most of the observed absorption resonances could be described using regular Lorentzian expressions for the complex dielectric permittivity

$$\varepsilon^*(\nu) = \varepsilon'(\nu) + i\varepsilon''(\nu) = \sum_j \frac{f_j}{\nu_j \gamma_j + i(\nu^2_j - \nu^2)}, \tag{3}$$

where $\varepsilon'(\nu) = n^2_2 - k^2_2$ and $\varepsilon''(\nu) = 2n_2 k_2$ are real and imaginary parts of $\varepsilon^*$, $f_j = \Delta\varepsilon_j\ \nu_j^2$ is the oscillator strength of the $j$-th resonance, $\Delta\varepsilon_j$ is its dielectric contribution, $\nu_j$ represents the resonance frequency and $\gamma_j$ the damping. For some resonance absorptions, however, we were not able to describe the shape of the curves in the $Tr(\nu)$ and $R(\nu)$ spectra with a Lorentzian (3).

Specifically, these are the absorption lines located at 10 cm$^{-1}$ – 30 cm$^{-1}$ and 120 cm$^{-1}$ – 160 cm$^{-1}$. Satisfactory descriptions could only be reached by using the expression for *coupled* Lorentzians proposed by Hopfield et al. for some dielectric compounds. The complex dielectric permittivity is then written as:[54]

$$\varepsilon(v) = \frac{f_1(v_2^2 - v^2 + iv\gamma_2) + s_2(v_1^2 - v^2 + iv\gamma_1) - 2\sqrt{f_1 f_2}(\alpha + iv\delta)}{(v_1^2 - v^2 + iv\gamma_1)(v_2^2 - v^2 + iv\gamma_2) - (\alpha + iv\delta)^2}, \qquad (4)$$

where $j = 1, 2$; $f_j = \Delta\varepsilon_j \, v_j^2$ is the oscillator strength of the $j$-th Lorentzian with $v_j$ being the eigenfrequency. $\alpha$ and $\delta$ are the real and the imaginary coupling constants, respectively. Successful fits by applying expression (4) indicate that the corresponding absorption processes are essentially not independent from each other.

Using least-square fits of the *Tr(v)* and *R(v)* spectra with the expressions (1) – (4) in combination with the conductivity and permittivity spectra measured directly at the terahertz-subterahertz range allow us to obtain broad-band spectra of real and imaginary parts of the dielectric permittivity and optical conductivity of the beryl crystal. Since the phonon absorption of beryl has been studied earlier (see, for example,[55, 56, 57]), here we will concentrate on the response caused exclusively by water molecules. To distinguish water-related absorptions from phonon resonances, we have performed comparative optical measurements of the samples before and after dehydration. In order to extract the crystal water from the samples they were heated to 1000 °C in vacuum for 24 hours leading to a weight loss of approximately 1%, corresponding to ~0.5 H$_2$O molecules per formula unit of Be$_3$Al$_2$Si$_6$O$_{18}$.

Characteristic transmission coefficient spectra of the sample before and after dehydration are presented in Fig. 2c. Disappearance of absorption structures in the dehydrated crystal is clearly seen pointing to their connection with the H$_2$O molecular response. It is important to note that all phonon resonances, like the minimum at 250 cm$^{-1}$ in Fig. 2c, stayed unchanged in the dehydrated crystals. Having identified the water-related absorption lines, the phonon resonances have been "subtracted" from the $\sigma^*(\nu) = \sigma_1(\nu) + i\sigma_2(\nu)$ and $\varepsilon^*(\nu) = \varepsilon'(\nu) + i\varepsilon''(\nu)$ spectra by setting corresponding oscillator strengths $f_j$ (expressions (3) and (4)) to zero. For presentation of the results we have chosen the spectra of the real parts of dielectric permittivity $\varepsilon'(\nu,T)$ and of the optical conductivity $\sigma_1(\nu,T) = \sigma(\nu,T) = \nu\cdot\varepsilon''(\nu,T)/2$. We display $\sigma(\nu,T)$ instead of $\varepsilon''(\nu,T)$ because the area under the resonance peaks seen in the $\sigma(\nu,T)$ spectra is directly connected to the oscillator strength of the resonance $f$

$$f = \frac{4}{\pi}\int \sigma(\nu)d\nu.$$

Broad-band optical conductivity spectra connected exclusively with the response of water within beryl crystal and measured for two polarizations at 5 K are presented in Fig. 3a. For comparison we display in the same figure spectra of the optical conductivity of liquid water and of the hexagonal ice, data taken from Refs.[58, 59, 60, 61]. Hatched parts in the figure denote the area where the phonon resonances were observed and subtracted; this procedure introduces some uncertainty in determination of the parameters of the water-related modes. The spectra shown in Fig. 3a are highly anisotropic, quantitatively and qualitatively, especially below ~1000 cm$^{-1}$, where absolute values of conductivity differ by one to three orders of magnitude. Also, whereas for the parallel $E\|c$ polarization at $\nu < 1000$ cm$^{-1}$ there are only two absorption lines at $\approx 90$ cm$^{-1}$ and $\approx 160$ cm$^{-1}$ much richer structures are observed for the $E\perp c$ case. Here three regions can be distinguished: a) a broad peak at the lowest frequency of $\approx 25$ cm$^{-1}$ with two narrower resonances at its high-frequency shoulder, b) two bands above 100 cm$^{-1}$ centered at $\approx 150$ cm$^{-1}$ and at $\approx 400$ cm$^{-1}$, each composed of several more narrow peaks (shown in details in Fig. 3b) and c) higher-frequencies ($\nu > 1000$ cm$^{-1}$) internal modes of the H$_2$O molecule. For both polarizations, $\boldsymbol{E}\perp c$ and $\boldsymbol{E}\|c$, peaks at $\approx 5300$ cm$^{-1}$ have been treated as combinations ($\nu_1 + \nu_2$) of the intramolecular $\nu_1$ and $\nu_2$ modes.

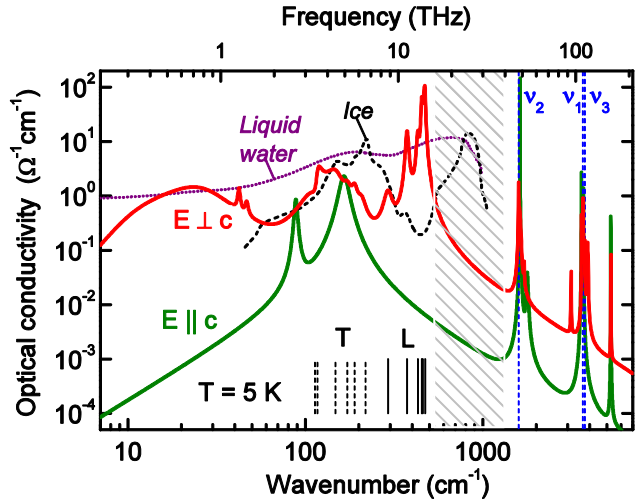

FIG. 3. (Color online). Spectra of optical conductivity caused by the response of water molecules in beryl nanocavities measured at T = 5 K for two principle polarizations, E⊥c and E∥c. The response of phonons (frequency range shown by hatching) was "subtracted" (see text). Dashed lines indicate frequencies of intramolecular stretching and bending modes ($\nu_1$, $\nu_2$, $\nu_3$) of the free water molecule. Doted (violet) and dashed (black) spectra correspond, respectively, to liquid water (T = 27°C, data from [58-60]) and hexagonal ice (T = 100 K, data from [61]). Transitions from the lowest-energy state (ground, G band, see below) to the translational band (T) and to the librational band (L) are indicated by dashed and solid lines respectively.

In Fig. 4 we show displacements of hydrogen and oxygen ions of the H$_2$O molecule which are associated with the intramolecular modes $\nu_1$, $\nu_2$, $\nu_3$ and the corresponding changes of the H$_2$O dipole moment. For a fixed polarization of the probing radiation, $\boldsymbol{E}\|c$ or $\boldsymbol{E}\perp c$, only two internal modes will be optically active for a separate water molecule, for example, $\nu_1$ and $\nu_2$ of water-II for $\boldsymbol{E}\|c$ or $\nu_1$ and $\nu_2$ of water-I for $\boldsymbol{E}\perp c$. Since for both polarizations all three internal modes are clearly seen (see Fig. 3a) we conclude that both types of water molecules are present in the studied sample.

## IV. DISCUSSION.

**Perpendicular polarization, E⊥c.**

In order to relate the observed absorption resonances to the dynamics of water molecules within the nano-cavities of beryl crystal lattice, we start from two assumptions:

1. There is noticeable coupling of the $H_2O$ molecules to the host crystal lattice via hydrogen bonds. Although this coupling might be relatively weak, there is an appreciable effect on the optical response of $H_2O$, at least at temperatures below 200 K, see.[37, 45] Our results indicate that this coupling can be observed even up to room temperatures.

2. Local conditions for the formation of hydrogen bonds between $H_2O$ molecules in liquid water or in ice are not essentially different from those forming the H-bonds between the $H_2O$ molecule and the "walls" of nano-cavities in beryl.

These assumptions imply the existence of translational (T) and librational (L) vibrations of water molecule in beryl nano-cavity; this is in analogy to the T- and L- bands in liquid water or in ice.[58 - 61] We also expect that the corresponding absorption bands in beryl should be located at roughly the same frequencies as those found in liquid water or in ice. Indeed, the similarity becomes evident from Fig. 3a between the infrared (100 cm$^{-1}$ to 1000 cm$^{-1}$) responses of the water molecules in beryl ($E \perp c$ case), on the one side, and liquid water and ice, on the other side. Unexpectedly, even the absolute values of optical conductivity in beryl and water/ice are roughly the same. The translational bands in liquid water and in ice are located at approximately the same frequency, around 200 cm$^{-1}$, while the librational band in ice is shifted to about 800 – 900 cm$^{-1}$ compared to the L-band in water observed at $\approx$ 600 cm$^{-1}$. Two separate infrared bands are distinctly recognized also in the conductivity spectrum of beryl. We associate the low-frequency band with the translations (oscillations) and the high-frequency band with the librations (restricted rotations) of the $H_2O$ molecule in a cavity of the beryl crystal lattice. In beryl, these T- and the L- bands are found at $\approx$ 140 cm$^{-1}$ and $\approx$ 400 cm$^{-1}$, respectively, noticeably lower than the corresponding bands in water and ice; one possible reason is the weaker bonding of the $H_2O$ molecules to the host lattice of beryl.

To account for the observed THz-IR water-related spectra (T- and L- bands, their fine structure and the lowest-frequency band around 20 cm$^{-1}$ – 40 cm$^{-1}$) we propose the following model.

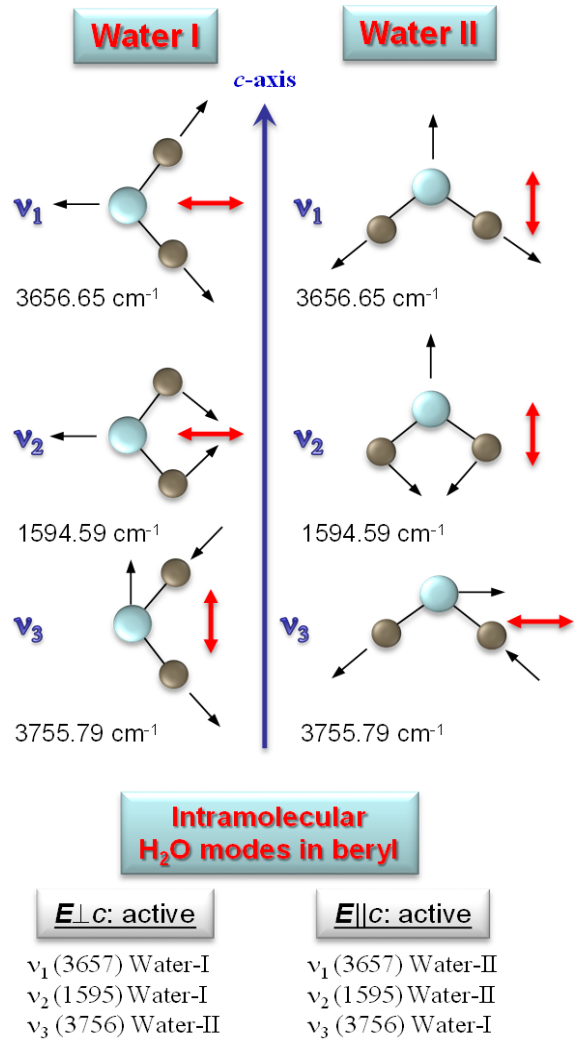

Fig.4. (Color online). Sketch showing the change of dipole moment of the water molecules (thick red arrows) caused by internal $H_2O$ vibrations $\nu_1$, $\nu_2$, $\nu_3$ (thin arrows indicate atomic shifts) together with the calculated values of eigenfrequencies (ex., [38]). Two orientations of water molecules relative to the crystallographic axis $c$ are drawn to illustrate the optical activity of internal vibrations of water-I and water-II molecules at different polarizations of the probing electromagnetic radiation: $E \| c$ and $E \perp c$.

**Model.**

Since we are considering the vibrations of water molecule at frequencies smaller than the intramolecular modes frequencies $v_1$, $v_2$, $v_3$, the molecule is treated as a rigid object characterized by its position and angular orientation. Correspondingly, its T-modes are related to excitation of positional vibrations and the L-modes - to excitation of rotational motions. Fig. 5 schematically shows the one-dimensional channels formed by the nano-cavities in the crystal structure and the two positions of the two types of water molecules, water-I and water-II, coupled to the crystalline surrounding via hydrogen bonds. The water-II molecules, in addition, are coupled to the cations by relatively stronger electrostatic forces. The shifts of the molecules corresponding to the T- and L- vibrations are shown by straight (T) and bent (L) arrows in Fig. 5a, b. The T-modes can be described in terms of the molecular displacements from the equilibrium position characterized by the vector *r(x,y,z)*. The corresponding Hamiltonian is

$$H_T = -\frac{\hbar^2}{2M}\Delta + U_T(x,y,z), \tag{5}$$

where $\hbar$ is the Planck's constant, $M$ is the molecule effective mass and $U$ is the energy related to the displacements relative to the crystalline environment. We assume that, in the first approximation, $U$ is a quadratic function that corresponds to harmonic oscillations of the molecule near the equilibrium position. Note that, in principle, the effective mass $M$ in Eq. (5) can differ from the mass of a free $H_2O$ molecule due to the coupling to the beryl crystalline environment. However, the deviations may be expected to be small due to the large difference of the masses of the $H_2O$ molecules and of the surrounding ions of the lattice.

It is clear that the frequencies of the molecular vibrations should be different for the three normal modes. For water-I, the modes are related to the displacements (a) along the dipole moment, (b) perpendicular to the dipole moment and to the *c*-axis, and (c) along the *c*-axis. For water-II, the modes are related to the displacements of the molecules (a) along the line passing through the hydrogen atoms, (b) perpendicular to this line and to the *c*-axis, and (c) along the *c*-axis. In addition to the translational degrees of freedom one should take into account three rotational degrees of freedom - three angles that describe the orientation of the water molecule within the crystalline cavity. Rotations of the water molecule in the cavity are not free. Within the six-well potential (sketched in Fig. 5c) there are preferred orientational positions that are different for water-I and water-II. Therefore, the orientational modes correspond to angular oscillations (librations) of the water molecule near these equilibrium orientations. There should be three corresponding librational modes. However, their assertion deserves an essential specification related to the hexagonal symmetry of the crystalline potential.

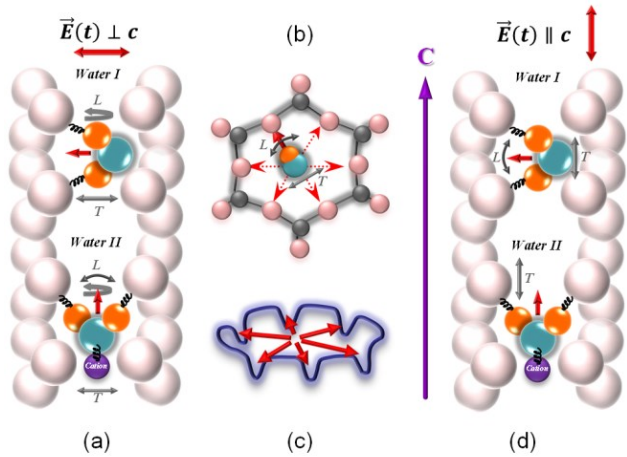

FIG. 5. (Color online). Schemes showing positions of water-I and water-II molecules in the cavities of beryl crystal lattice. Thick red arrows mark the dipole moments of $H_2O$. Thin and thick "springs" denote hydrogen bonding of $H_2O$ to the host crystal framework and its electrostatic bonding to the cation, respectively. Straight and bent grey arrows indicate translational (T) and librational (L) external vibrations of the $H_2O$ molecules subject to the electric field vector $E(t)$ perpendicular (a,b,c) or parallel (d) to the crystallographic axis *c*. Panel (b) shows a view at the water-I molecule along the *c*-axis. Panel (c) shows a six-well potential sensed by water molecules when rotated around the *c*-axis.

We first consider the response of the water-I molecule. Its rotations around the *c*-axis, viz. its librational modes related to these rotations, are described by the Hamiltonian:

$$H_L = -\frac{\hbar^2}{2I}\frac{\partial^2}{\partial \phi^2} + U_L(\phi), \tag{6}$$

where $I$ is the moment of inertia of the molecule, $\varphi$ the rotational angle around the $c$-axis and $U_L$ the energy related to the rotations of the molecule relative to the crystalline environment. Due to the hexagonal symmetry we have $U_L(\varphi) = U_L(\varphi+2\pi/6)$, implying six preferred directions of the molecule orientation where the potential $U_L(\varphi)$ attains minima. The assumption that the H$_2$O molecule is weakly H-bonded to the surrounding implies that quantum tunneling becomes relevant leading to a hybridization of the molecular states associated to the separate minima. The resultant hybrid states can be characterized by the angular number $m = 0, \ldots, 5$. The wave function $\Psi$ can be written as:

$$\Psi(\phi + \frac{\pi}{3}) = \exp(\frac{im\pi}{3})\Psi(\phi). \tag{7}$$

The above reasoning deals with the quantum states that are generated by hybridization of the states localized near the minima of the potential $U_L(\varphi)$. These states can be ground states, or can be excited states that are related to translational and/or librational motion of the molecule. Therefore, any molecular state can be characterized by six quantum numbers corresponding to excitation of three translational modes and three librational modes. One expects that energy shifts related to quantum tunneling are relatively weak. Consequently, the molecular states are split into sets with close energies characterized by different numbers m but identical other quantum numbers. We shall call such sets as bands.

There exists an additional symmetry of the beryl nano-cavity that is invariant under reflections with respect to certain planes passing through the $c$-axis. This requires $U_L(\varphi) = U_L(-\varphi)$ if the preferred direction lies in the symmetry plane. Then the states belonging to a band will be degenerate: energies of the states with $m = 1, 5$ are the same and those with $m = 2, 4$ are also the same. It is explained by the fact that the transformation $\psi(\phi) \to \psi(-\phi)$ is equivalent to the transformation $m \to 6 - m$. Strictly speaking, the degeneracy has to be observed for the band originating from the ground state. However, one expects that this property is approximately correct for bands originating from excited states, as well.

For the optical spectroscopy there are selection rules for the transitions related to the character of the interaction of the electromagnetic field with the water-I and water-II molecules in beryl. In a first approximation the dipole interaction energy $-D \cdot E$ for water-I molecule can be written as

$$H_{int} = -D_0(E_x \cos\phi + E_y \sin\phi), \tag{8}$$

where $D_0$ is the molecule dipole moment, and $E_x$, $E_y$ are the components of the radiation electric field in the plane perpendicular to the *c*-axis. Equation (8) implies that within the band only transitions with $\Delta m = \pm 1$ are allowed (keeping $m = 6$ equivalent to $m = 0$).

According to Hamiltonian (8) transitions between different bands are not allowed due to orthogonality of the corresponding quantum states. However, there are corrections to the dipole moment of the system related to the polarization of the water molecule environment. These corrections can be included into expression (8); they are sensitive to the position and to the orientation of the water molecule. Taking these corrections into account transitions between different bands become allowed. Consequently, the optical activity of the T and L vibrations appears due to local fields that lead to a nonzero dipole moment of the complex "$H_2O$+crystalline surrounding".[62, 63] In any case, for the electric field $E \perp c$ the corrections have the same transformation properties with respect to the angle $\varphi$ as the expression (8). Therefore the selection rule $\Delta m = \pm 1$ holds for the interband transitions as well.

Now we discuss the response of water-II molecule. Its rotations around the c-axis can be treated in analogy with the analysis done for water-I, with the similar Hamiltonian given by expression (6). The quantum states of water-II molecule can also be characterized by the angular number $m$. However, in this case the molecule returns to the same state after a rotation by the angle $\pi$ (instead of $2\pi$) because of the identity of the hydrogen atoms. For the wavefunctions we obtain the condition $\Psi(\varphi+\pi) = \Psi(\varphi)$, that leads to the restriction of quantum numbers $m = 0, 1, 2$ for the molecular quantum states. The states with $m = 1, 2$ are (approximately) degenerate. Since the dipole moment of water-II molecule is oriented along the $c$-axis, for the polarization $\boldsymbol{E} \perp c$ interband transitions are also possible with the selection rule $\Delta m = \pm 1$.

**Assignment of absorption lines.**

We apply the above considerations to interpret the water-related absorption lines observed in the spectra of beryl. In the $\boldsymbol{E} \perp c$ optical conductivity spectrum shown in Fig. 3 we reliably detect six absorption peaks in the T-band and six peaks in the L-band. We cannot exclude the possibility of existence of more resonances that should originate from different types of T and L modes as described above; these additional modes might be masked by strong phonon features. Since water-I molecules are more loosely coupled (via H-bonds) to the surrounding ions than the water-II molecules (electrostatic interaction with cations), we believe that the rich set of absorption lines seen in the spectra for $\boldsymbol{E} \perp c$ (Fig. 3) originates from the response of the water-I molecules. Figure 6 depicts a simplified scheme of energy levels of the water-I molecules ground (G) band and translational (T) and librational (L) bands corresponding to one separate type of T and L modes, in accordance with the above model. In the figure the allowed intra- and interband transitions are marked by arrows. We associate these transitions with the resonances in the spectra of Fig. 3. The eigenfrequencies and other parameters of the transitions from the G-band to the T-band and to the L-band are summarized in Table I. We indicate corresponding energies of the transitions in meV and frequencies in cm$^{-1}$ (Fig. 6c, d). Two dashed arrows depicting the G→T transitions in Fig. 6c correspond to the transitions to the levels with $m = 0$ and $m = 3$, whose location in the conductivity spectra could only be determined with a relatively large uncertainty caused by the procedure of extracting of water-related response mixed with phonon resonances, as described above. The dotted arrows in Fig. 6c, d mark the allowed intraband transitions within the T- and L-bands. The intervals between the corresponding energy levels were estimated according to the known "lengths" of different arrows that indicate the inter-band (G to L and G to T) transitions. The significant difference between the obtained values deserve further study, but can in part be explained by the uncertainty in determination of the eigenfrequencies of the transitions shown by dashed arrows. We did not identify the intraband absorptions (dotted arrows) in our experimental spectra possibly because of their relatively low intensities (low energy levels population) or overdamped character.

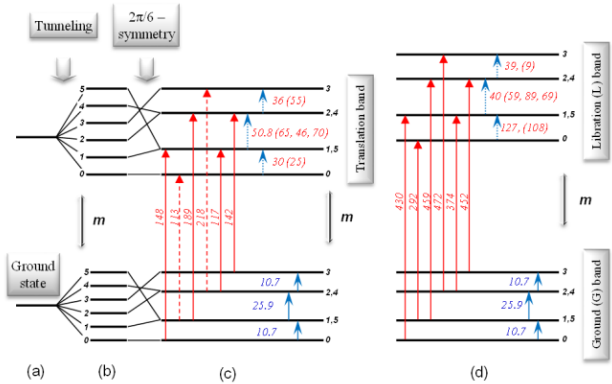

FIG.6. (Color online). Scheme of vibrational energy levels of water-I molecule in nano-cavity of beryl crystal lattice. Two lowest levels are shown. (a) Ground state and the first excited state are split into six sublevels due to tunneling (b), angular quantum number m, in the six-well potential with levels $m = 1$, $m = 5$ and $m = 2$, $m = 4$ degenerate in couples (c) due to six-fold symmetry of the crystalline potential. Long (red) and short (blue) arrows indicate inter- and intra-band optical transitions, respectively, allowed by the selection rule $\Delta m = \pm 1$. Dashed arrows indicate transitions whose energies were determined with additional uncertainty due to phonon absorptions present in the correspondent frequency range. Dotted arrows correspond to transitions not seen in the experiment. Frequencies (in cm$^{-1}$) of the transitions are indicated, see text.

Within our model, we ascribe the wide ridge near 25 cm$^{-1}$ in the conductivity spectrum (see Fig. 3) to transitions between the states within the G-band. Its complicated spectral shape (see Fig. 6c, d) was reproduced with coupled Lorentzians according to Equation (4) and can be attributed to cross populations of intra-ground band levels numbered with $m = 0, 1, …, 5$. Additionally, the spectral shape of the 25 cm$^{-1}$ band (as well as of the other absorptions) can be influenced by inhomogeneous broadening of the G-band levels that is related to the electric field of charged impurities (metallic ions) present in the crystal. The electric field randomly shifts the energy levels of the G-band due to interaction with the water molecule via its dipole moment. A detailed analysis of this kind of inhomogeneous broadening is a subject of special investigation and is beyond of the scope of the present paper.

TABLE I. Parameters and assignments of water-related modes observed in beryl for polarization $E \perp c$ at $T = 5$ K: $E$ – energy, $v_0$ – eigenfrequency, $f$ – oscillator strength (intensity), $\gamma$ – damping. Stars mark the values obtained using the model of coupled Lorentzians (expression 4).

| **Terahertz band** | | | | | |
|---|---|---|---|---|---|
| E (meV) | $v_0$ (cm$^{-1}$) | f (cm$^{-2}$) | $\gamma$ (cm$^{-1}$) | (G,m)-(G,m) | Assignment |
| 1.33 | 10.7* | 106 * | 12 * | (G,0)-(G,1) | Water-I. Intra-ground (G) band |
| 3.21 | 25.9* | 1 770 * | 25 * | (G,1)-(G,2) | Water-I. Intra-ground (G) band |
| 5.21 | 42 | 74 | 1.5 | ? | Water-II |
| 5.82 | 47 | 134 | 6 | ? | Water-II |
| **Far-infrared translation band** | | | | | |
| E (meV) | $v_0$ (cm$^{-1}$) | f (cm$^{-2}$) | $\gamma$ (cm$^{-1}$) | (G,m)-(T,m) | Assignment |
| 14.01 | 113 | 3 320 | 43 | (G,1)-(T,0) | |
| 14.51 | 117 * | 2 960 * | 8 * | (G,2)-(T,1) | Water-I. T-band: transitions between ground band (G) and translation band (T) |
| 18.35 | 148 * | 1 860 * | 34 * | (G,0)-(T,1) | |
| 21.33 | 172 | 900 | 19 | (G,3)-(T,2) | |
| 23.44 | 189 | 780 | 15 | (G,1)-(T,2) | |
| 27 | 218 | 1300 | 42 | (G,2)-(T,3) | |
| **Far-infrared libration band** | | | | | |
| E (meV) | $v_0$ (cm$^{-1}$) | f (cm$^{-2}$) | $\gamma$ (cm$^{-1}$) | (G,m)-(L,m) | Assignment |
| 36.21 | 292 | 2 020 | 34 | (G,1)-(L,0) | |
| 46.38 | 374 | 13 220 | 15 | (G,2)-(L,1) | Water-I. L-band: transitions between ground band (G) and libration band (L) |
| 53.32 | 430 | 10 750 | 14 | (G,0)-(L,1) | |
| 56.05 | 452 | 26 600 | 8 | (G,3)-(L,2) | |
| 56.92 | 459 | 10 110 | 5.9 | (G,1)-(L,2) | |
| 58.53 | 472 | 42 870 | 9 | (G,2)-(L,3) | |

In addition to the 25 cm$^{-1}$ bump, there are two rather narrow absorption lines observed for $E \perp c$ between 40 and 50 cm$^{-1}$ up to 200 K – 300 K. They do not fall into a pattern of water-I vibrations as described above. Their nature can be assigned to the response of water-II molecules, whose stronger coupling to the cations (compared to type-I molecules that are H-bonded to the cavity walls) leads to significantly different characteristics of the resonances, namely, smaller damping and spectral weight.

Table II summarizes the parameters of water-related absorption lines in beryl observed for $E \perp c$ in the infrared range (Fig. 3a) around $\nu_1$, $\nu_2$ and $\nu_3$ intramolecular modes. These resonances are presented in detail in Fig. 7. We consider the infrared satellites as combinations of $\nu_1$, $\nu_2$, $\nu_3$ modes with lower frequency excitations, whose locations should then correspond to the differences between the peak positions in Fig. 7. The difference frequencies are also presented in Table II. It can be seen that the values of the difference frequencies mostly fall below ≈200 cm$^{-1}$ matching well with the peaks positions related to the transitions from the G-band to the T-band. This agreement suggests that the intramolecular modes of H$_2$O molecule most easily couple to its translational vibrations.[44] At the same time, there are only a few rather intense infrared resonances that produce difference frequencies above 200 cm$^{-1}$ – 300 cm$^{-1}$, up to ≈530 cm$^{-1}$. They correspond to excitations to the librational band and suggest some degree of coupling of H$_2$O internal modes to the rotational degrees of freedom as well. For several infrared absorptions there are no counterparts in the far-infrared or THz-subTHz ranges, presumably due to their low intensity (small spectral weight $f$) and/or large damping.

**Parallel polarization, E∥c.**

In contrast to the $E \perp c$ polarization, the experimental water-related spectra for the polarization $E \| c$ are very simple: at 5 K only two resonances are seen at far-infrared and terahertz frequencies (88 cm$^{-1}$ and 158 cm$^{-1}$, see Fig. 3a), and just two satellite peaks accompany the $H_2O$ intramolecular modes (see Fig. 7) with their parameters given in Table III. We assign the peaks at 158 cm$^{-1}$ and at 88 cm$^{-1}$ to water-I translations and librations, respectively.

**Temperature behavior.**

The overall temperature behavior of water-related absorption lines (except the $H_2O$ intramolecular modes) in beryl for both polarizations is illustrated in Fig. 8 where the spectra of the optical conductivity are presented for two temperatures, 5 K and 300 K.

TABLE II Parameters and assignments of water-related modes observed in beryl for polarization $E \perp c$ at $T = 5$ K: $E$ – energy, $v_0$ – eigenfrequency, $f$ – oscillator strength (intensity), $\gamma$ – damping. Bold numbers correspond to intramolecular $H_2O$ modes. Differences between positions of $v_1$, $v_2$ and $v_3$ intramolecular water vibrations and of the observed modes are presented.

| Infrared modes around internal $v_2$ (1595 cm$^{-1}$) mode of free $H_2O$ molecule | | | | |
|---|---|---|---|---|
| E (meV) | $v_0$ (cm$^{-1}$) | f (cm$^{-2}$) | $\gamma$ (cm$^{-1}$) | $\|v_{2\text{water-I}} - v_0\|$ (cm$^{-1}$) |
| **197.3** | **1 591** | **695** | **7.4** | $v_2$, Water-I |
| 197.7 | 1 594 | 84 | 2.6 | 3 |
| 198.0 | 1 597 | 215 | 5 | 6 |
| 199.9 | 1 612 | 857 | 29 | 21 |
| 202.7 | 1 634 | 190 | 25 | 43 |
| 212.0 | 1 710 | 29 | 12 | 119 |

| Infrared modes around internal $v_1$ (3657 cm$^{-1}$) and $v_3$ (3756 cm$^{-1}$) modes of free $H_2O$ molecule | | | | | |
|---|---|---|---|---|---|
| E (meV) | $v_0$ (cm$^{-1}$) | f (cm$^{-2}$) | $\gamma$ (cm$^{-1}$) | $\|v_{1\text{water-I}} - v_0\|$ (cm$^{-1}$) | $\|v_{3\text{water-II}} - v_0\|$ (cm$^{-1}$) |
| 389.9 | 3 144 | 32 | 14 | 461 | 528 |
| 446.2 | 3 598 | 80 | 12 | 7 | 74 |
| **446.9** | **3 604** | **277** | **0.8** | $v_1$, Water-I | |
| 449.5 | 3 625 | 70 | 52 | 20 | 47 |
| 453.7 | 3 659 | 57 | 4 | 54 | 13 |
| 454.3 | 3 664 | 270 | 8 | 59 | 8 |
| 454.8 | 3 668 | 70 | 4.4 | 63 | 4 |
| 455.1 | 3 670 | 15 | 1.8 | 65 | 2 |
| **455.3** | **3 672** | **342** | **6.6** | | $v_3$, Water-II |
| 455.7 | 3 675 | 36 | 3 | 71 | 3 |
| 456.0 | 3 677 | 225 | 5.5 | 73 | 5 |
| 457.1 | 3 686 | 136 | 5 | 82 | 14 |
| 458.7 | 3 699 | 20 | 3 | 95 | 27 |
| 461.8 | 3 724 | 144 | 20 | 120 | 52 |
| 463.9 | 3 741 | 4 | 7 | 137 | 69 |
| 465.4 | 3 753 | 183 | 22 | 149 | 81 |
| 479.4 | 3 866 | 40 | 13 | 262 | 194 |
| 483.0 | 3 895 | 250 | 30 | 291 | 223 |
| 486.1 | 3 920 | 60 | 15 | 316 | 248 |
| 653.5 | 5 270 | 50 | 7 | 1 665 | 1 598 |

We first consider the perpendicular polarization, $E \perp c$. It is seen that at 300 K all translational resonances are shifted to higher frequencies relative to their positions at 5 K. One expects that with increasing temperature the potential relief felt by the water molecule is partly smoothed due to thermal fluctuations. Therefore, the tunnel exponent characterizing the states within the band increases. That leads to an increase of the energy differences between the levels within the bands and to a shift of the peaks to higher frequencies. At the same time, the positions of the transitions to the L-band are practically temperature independent. This observation indicates that the librational motions of the $H_2O$ molecule are less sensitive to the thermal phonon bath. The detailed temperature variation of all modes parameters of the T- and L-bands are presented in Figures A1 - A4 of the Appendix.

TABLE III. Parameters and assignments of water-related modes observed in beryl for polarization $E \| c$ at $T = 5$ K: $E$ – energy, $v_0$ – eigenfrequency, $f$ – oscillator strength (intensity), $\gamma$ – damping. Bold numbers show parameters of intramolecular $H_2O$ modes. For infrared range, differences are presented between positions of $v_1$, $v_2$ and $v_3$ intramolecular water vibrations and of the observed modes.

| **Terahertz band** | | | | |
|---|---|---|---|---|
| E (meV) | $v_0$ (cm$^{-1}$) | f (cm$^{-2}$) | $\gamma$ (cm$^{-1}$) | Assignment |
| 10.91 | 88 | 200 | 4 | Translation mode of Water-I |
| 19.6 | 158 | 2 780 | 22 | Librational mode of Water-I |
| **Infrared modes around $v_2$ (1595 cm$^{-1}$) of free $H_2O$ molecule** | | | | |
| E (meV) | $v_0$ (cm$^{-1}$) | f (cm$^{-2}$) | $\gamma$ (cm$^{-1}$) | $|v_{2\text{water-II}} - v_0|$ (cm$^{-1}$) |
| 197.2 | 1 590 | 21 | 1.7 | 32 |
| 201.1 | **1 622** | **12 450** | **0.7** | **$v_2$, Water-II** |
| 220.7 | 1 780 | 82 | 37 | 158 |
| **Infrared modes around $v_1$ (3657 cm$^{-1}$) and $v_3$ (3756 cm$^{-1}$) of free molecule mode** | | | | |
| E (meV) | $v_0$ (cm$^{-1}$) | f (cm$^{-2}$) | $\gamma$ (cm$^{-1}$) | Assignment |
| 446.3 | **3 599** | **955** | **5.5** | **$v_1$, Water-II** |
| 458.3 | **3 696** | **490** | **5.6** | **$v_3$, Water-I** |
| 653.4 | 5 269 | 95 | 2.9 | $v_1 + v_2$ (= 5 221 cm$^{-1}$), Water-II |

We finally discuss the quite complicated lowest-energy dynamics revealed by the water-related spectra collected at terahertz frequencies, below 100 cm$^{-1}$ (see Fig. 9). With increasing temperature the broad 25 cm$^{-1}$ bump moves to higher frequencies and looses intensity. This is confirmed by the data presented in Fig. 10a, b where the temperature dependence of the parameters used to describe the spectral shape of the 25 cm$^{-1}$ bump by two coupled Lorentzians (Equation (4)) are shown. The monotonous decrease of the combined oscillator strength (open symbols in Fig. 10b) is clearly seen and can be ascribed to the temperature driven depopulation of the states of the G-band (Fig. 6). At the same time the intensities $f_j$ of the separate components are markedly different. Below ≈70 K the oscillator strength of the lower-frequency component (transitions from level 0 to levels 1 and 5 and from 2 and 4 to 3) decreases while the intensity of the lower-frequency component (transitions from level 0 to levels 1 and 5 and from 2 and 4 to 3) increases. An opposite trend is, however, expected if one assumes that the energy levels population is governed by the corresponding Boltzmann exponents. Quite different is also the behavior of the damping constants of the two components: while cooling, the low-frequency component becomes narrower whereas the high-frequency component broadens, most pronounced below ≈70 K (see Fig. 10c). In the same temperature range a change from independent (vanishing coupling constant, see Fig. 10e) to coupled dynamics occurs of the two components. Taking all these observations into account it becomes obvious that the dynamics of the $H_2O$ molecule is more complex than schematically presented by the energy levels scheme in Fig. 6.

As can be seen from Fig. 10 and also Figs. A1 - A4 of the Appendix, some parameters of the modes display an anomalous behavior at temperatures between 50 and 100 K. The anomalies appear only for the perpendicular polarization $E \perp c$. Anomalous behavior has also been observed below 150 K in quasielastic neutron scattering experiments on beryl crystal in Ref.[64] where it was assigned to freezing out of the diffusion of water molecules along the channel (along the $c$-axis). It is remarkable, that at the same temperatures the maximum is observed of the difference in the heat capacities of a beryl crystal with and without water, as demonstrated by Fig. A6. In Ref.[45] an abrupt decrease of the $v_1$ intramolecular mode frequency and of the intensity ratios of intramolecular modes $v_3$ to $v_1$ were found in roughly the same temperature interval (55 K) and ascribed to the shift of the $H_2O$ molecule towards the crystal lattice cavity wall that occurs at low temperatures. This could be one reason for the anomalous behavior we observe. Another possibility could be that by lowering the temperature an interaction between neighboring water molecules comes into play leading to certain ordered arrangement of the $H_2O$ molecules in the channels.

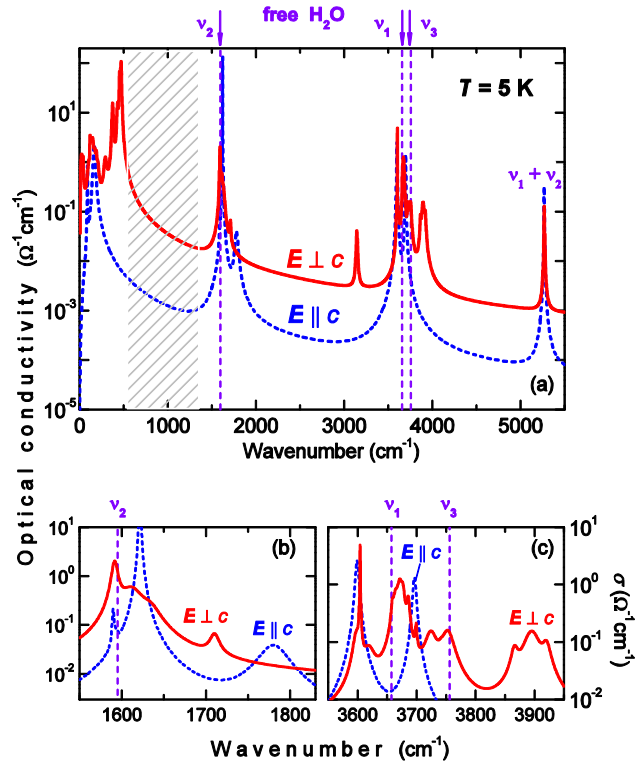

FIG.7. (Color online). (a) Spectra of the optical conductivity caused by the response of water molecules in beryl nano-cavities measured at T = 5 K for polarizations E⊥c (solid, red) and E∥c (dashed, blue). Arrows and dashed lines indicate the frequencies of free $H_2O$ intramolecular stretching and bending modes $v_1$, $v_2$, $v_3$. Panels (b) and (c) display in details the satellite absorption peaks around $v_1$, $v_2$, $v_3$.

From Fig. 9 it is seen that during heating up above 220 K, the oscillator strength of the 25 cm$^{-1}$ bump becomes much smaller, but in addition another mode between 50 cm$^{-1}$ – 60 cm$^{-1}$ grows out and finally at 290 K forms a broad peak centered at about 56 cm$^{-1}$. A very similar resonance absorption between 50 cm$^{-1}$ – 60 cm$^{-1}$ has been observed in liquid water, by infrared spectra, Raman scattering and neutron experiments.[65, 66] A detailed review of experimental observations and discussions of the origin of this so-called 60 cm$^{-1}$ mode is given in.[67, 68] Experimentally, the 60 cm$^{-1}$ mode is hardly observable by infrared spectroscopy due to large background absorption of water (see, for example, [69, 70]) but is clearly detected in the Raman measurements (ex., [71, 72]) and well reproduced by simulations.[67, 73, 74] A 60 cm$^{-1}$ mode, however, is reliably detected in the infrared spectra of ice (see, ex., [61]]). Although the nature of the 60 cm$^{-1}$ mode in water and in ice is frequently ascribed to vibrations of the H$_2$O molecule that involve bending of the hydrogen bonds binding it to other water molecules, serious doubts on this interpretation have been put forward recently[73, 74, 75, 76] showing that the origin of the mode is still under debate.

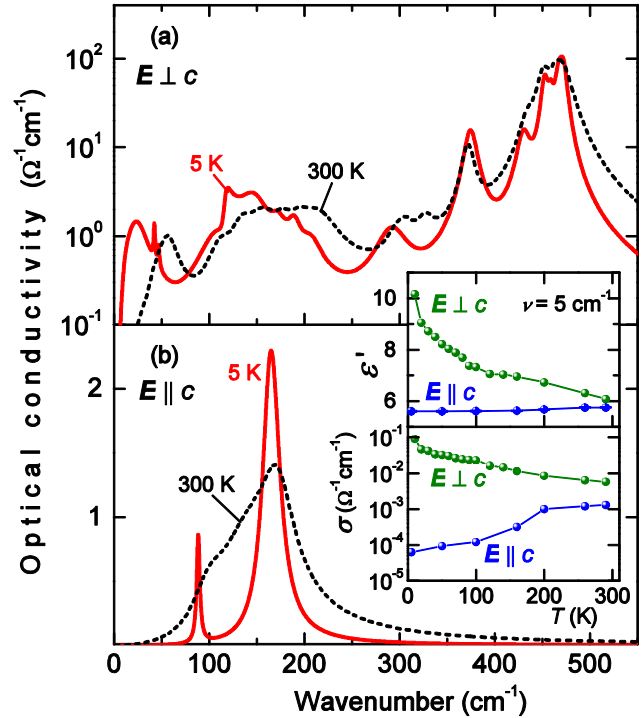

FIG.8. (Color online). Spectra of the optical conductivity caused by the response of water molecules in beryl nano-cavities measured for polarizations E⊥c (a) and E∥c (b) at T = 5 K (red solid lines) and at T = 300 K (black dashed lines). The inset shows the temperature dependence of the permittivity and conductivity measured at low frequency of 5 cm$^{-1}$.

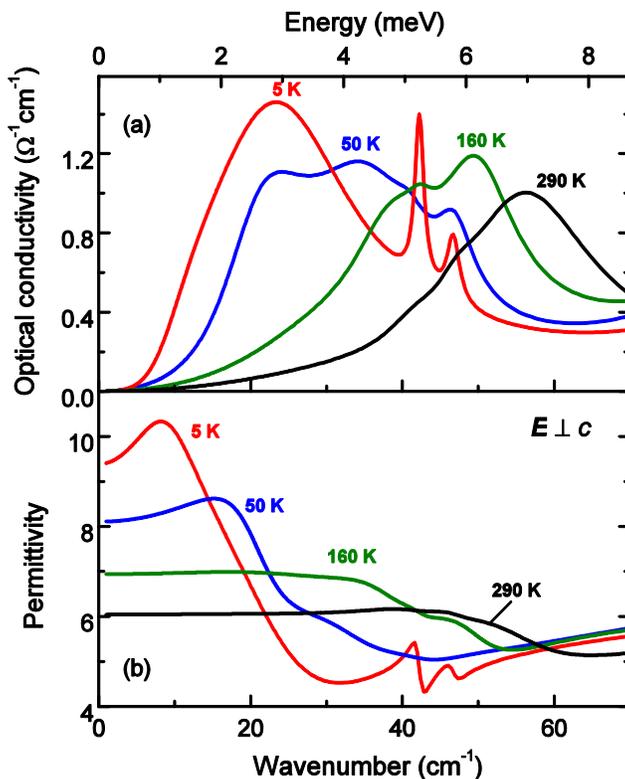

FIG.9. (Color online) Terahertz spectra of the optical conductivity and dielectric permittivity caused by the response of water molecules in beryl nano-cavities measured for polarization E⊥c at different temperatures, as indicated.

We assume that the resonance we observe in beryl at 56 cm$^{-1}$ at 290 K could have the same origin as the 60 cm$^{-1}$ mode in liquid water or in ice. Compared to liquid water or ice, in the spectra of beryl (Figs. 8, 9) the mode is very well pronounced. This can be a consequence of the fact that the water molecules in beryl are within a highly symmetric surrounding formed by its crystal lattice ions with which the $H_2O$ molecules interact. As for now, we cannot decide about the microscopic origin of this high-temperature excitation, whether it is determined by hydrogen bonds or by some sort of a cage effects due to dipoles induced by the intermolecular fields. However, our results indicate two important features: a) this mode bares essentially local character in agreement with experiments on aqueous solutions [72, 77, 78, 79] and with simulations;[74, 80] this basically means that it does not involve large complexes of $H_2O$ molecules. Moreover, since the beryl cavities contain just a single $H_2O$, our data evidence that the mode is of extreme local – single molecular character; b) it is important to note the extremely anisotropic nature of the mode in beryl: no signs of it are observed in the polarization $E\|c$, i.e. it is excited only when the electric field vector of the probing radiation is perpendicular to the line connecting two protons that would speak in favor of the H-bond bending origin of the mode.

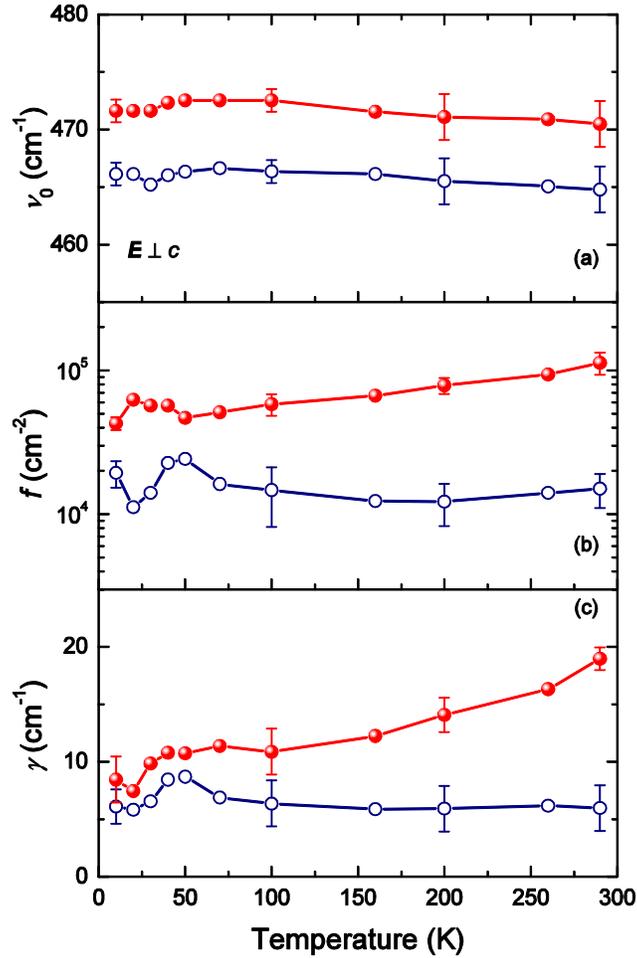

Fig.10. (Color online) Temperature dependences of the parameters of the water-related absorption lines in beryl observed at terahertz frequencies for polarization the E⊥c: eigenfrequencies $\nu_0$ (a, e), oscillator strengths f (b, f), damping factors γ (c, g) and coupling constant (d), cf. Eq. (4) in the text.

Finally, we note that the temperature behavior of the modes for the parallel, $E\|c$ polarization is rather featureless. Here, the two infrared absorption lines (88 cm$^{-1}$ and 158 cm$^{-1}$) do not reveal any unusual behavior, as seen in Fig. 8b. Their damping parameters decrease with cooling down that can be attributed to the phonon assisted broadening (Fig. A5). The intensity of the lower-frequency mode decreases towards low temperatures; some change is displayed by the strength of the higher-frequency peak. Slight decrease while cooling is shown by eigenfrequencies of both resonances.

## V. CONCLUSIONS.

We have presented a comprehensive study of the terahertz-infrared spectra of conductivity σ(ν) and dielectric permittivity ε'(ν) of beryl (morganite, Mn:Be$_3$Al$_2$Si$_6$O$_{18}$) measured in a broad range, from sub-terahertz frequencies of few wavenumbers up to the infrared, ν = 7000 cm$^{-1}$, at temperatures from 5 K to 300 K and for two polarizations relative to the crystallographic axis c: ***E***||*c* and ***E***⊥*c*. By comparing the spectra of as-grown crystals and of dehydrated (annealed) samples, we extract the σ(ν) and ε'(ν) spectra connected exclusively with dynamics of water molecules, which are captured individually within the nano-cavities of beryl crystal lattice and have their dipole moments aligned perpendicular (water-I) and parallel (water-II) to the c-axis. We find that the water-related spectra display highly anisotropic resonance structures, especially at far-infrared and terahertz frequencies. Here, only two absorption peaks (at ~90 cm$^{-1}$ and ~160 cm$^{-1}$) are seen for ***E***||*c*, while a much richer spectral pattern is recorded for the polarization ***E***⊥*c*. At liquid helium temperatures, for ***E***⊥*c* we observe two broad infrared bands around 150 cm$^{-1}$ – 200 cm$^{-1}$ and around 400 cm$^{-1}$ – 500 cm$^{-1}$, each consisting of a number of narrower peaks, and a terahertz bump centered at 20 cm$^{-1}$ – 30 cm$^{-1}$ with two sharp resonances at its high-frequency shoulder. Based on the analysis of our spectra in combination with the literature data on infrared response of liquid water and of ice, we associate the low and the high frequency infrared bands with the translational (T) and librational (L) vibrations of water-I molecules, respectively. We propose symmetry arguments that consider dynamical properties of H$_2$O molecule which experience a six-well parabolic potential formed by the crystalline surrounding. Within our model, the fine structure of the T and L bands is determined by optically active transitions from the molecular ground state to the corresponding first excited states each split due to quantum tunneling between the states in the potential relief minima. The broad bump in the terahertz range is assigned to transitions within the ground state "multiplet". The quantitative analysis of the spectra yields the temperature dependences of all water-related absorption lines parameters: resonance frequencies, oscillator strengths and damping constants. These dependences together with the absolute values of the parameters indicate that the assumption about parabolic shape of the potential minima is oversimplified and that additional factors, like influence of impurities or interaction between neighboring water molecules, should be taken into account. The obtained results may help to analyze more complicated systems with confined water such as $H_2O$ chains in carbon nanotubes, molecular clusters in zeolites, clays, silica gels, and other natural or synthetic frameworks, and interfacial water in biological systems.

**ACKNOWLEDGEMENTS.**

Authors acknowledge fruitful discussions with B. Gompf, K. Lassmann, L.S. Yaguzhinskii, K.A. Motovilov. We thank Dan Wu, N. Aksenov and G. Chanda for their help with the infrared measurements, G. Untereiner for samples preparation, G. Siegle and E. Brücher for expert experimental assistance. The research was supported by the Russian Foundation for Basic Research project 14-02-00255-a, by the Russian Academy of Science Program for fundamental research "Problems of Radiophysics" and by the Ministry of Education and Science of the Russian Federation.

**APPENDIX: TEMPERATURE BEHAVIOR OF WATER RELATED RESONANCES**

In Figs. A1 - A4 we present in detail the temperature dependences of the parameters obtained from the water-related absorption resonances which correspond to transitions from the ground band to the translation and libration bands for the polarization $E \perp c$. In addition, Fig. A5 shows the similar dependences for the other polarization, $E\|c$. Only weak changes of the damping constants are observed for these essential modes in the broad range of temperature from 290 K to 100 K; this implies that the damping mechanism is mostly determined by impurities.

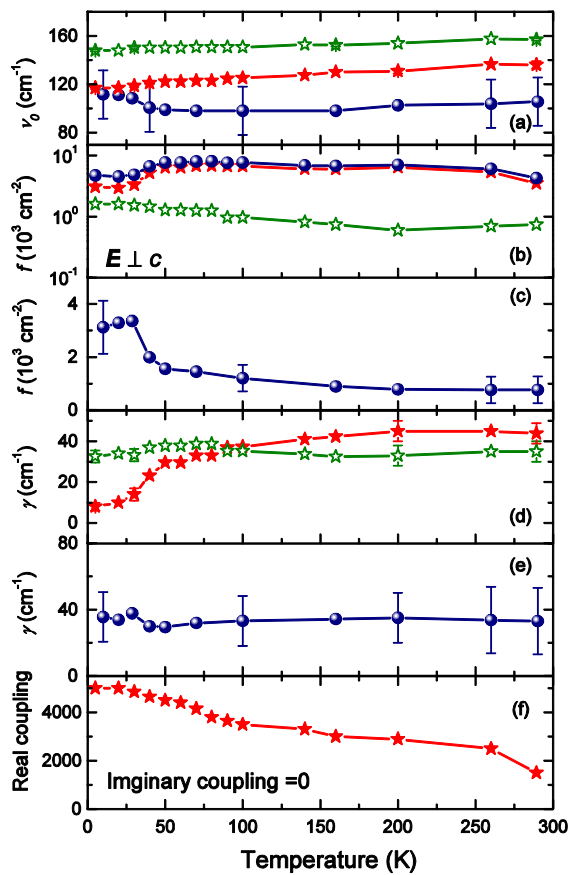

FIG. A1. (Color online). Detailed temperature dependences of parameters of the water-related absorption lines in beryl observed in the far-infrared range for polarizations $E\perp c$: eigenfrequencies $\nu_0$, oscillator strengths f, damping factors $\gamma$ and coupling constant.

Fig. A6 displays the heat capacities, in $C_p(T)/T$ and $\Delta C_p(T)/T$ representations, of a beryl crystal and of a crystal measured after depletion of the crystal water, together with the literature data collected on a crystal with 0.36 $H_2O$ per formula unit.[81] There is a clear difference between the data obtained on the water-containing and the dehydrated samples. The difference between the two data sets reaches its maximum value around 40 K – 50 K, i.e. at temperatures where oscillator strengths and dampings of some optical resonances reveal anomalous behavior, see main text. The dependence $\Delta C_p(T)/T$ was fitted to the sum of the heat capacities of four Bose-Einstein oscillators according to

$$C_V(T) = 3R \sum_{i=1}^{4} W_i \left(\frac{E_i}{k_B T}\right)^2 \frac{\exp(E_i/k_B T)}{\left[\exp(E_i/k_B T) - 1\right]^2} , \qquad (A1)$$

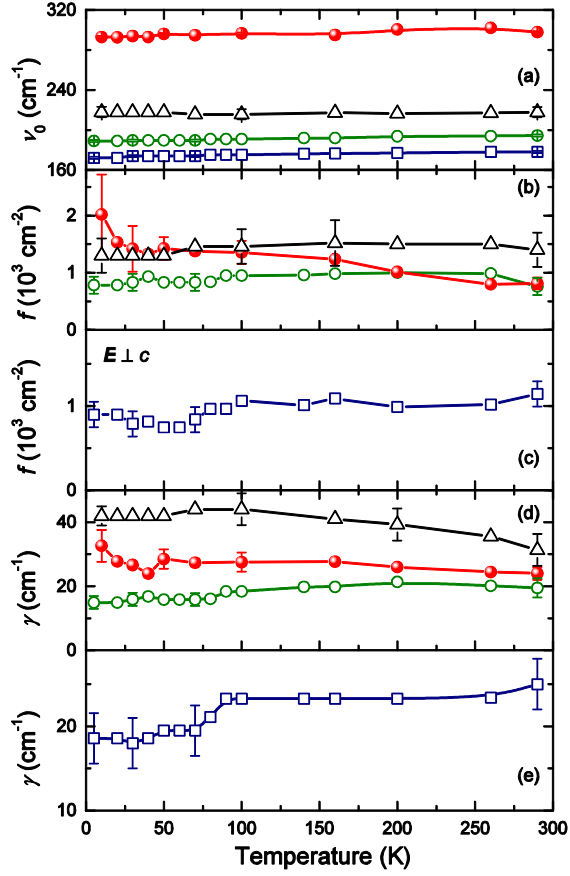

FIG. A2. (Color online). Detailed temperature dependences of parameters of the water-related absorption lines in beryl observed in the far-infrared range for polarizations $\mathbf{E} \perp c$: eigenfrequencies $\nu_0$, oscillator strengths $f$, damping factors $\gamma$ and coupling constant.

where $E_i$ are the energies of the individual oscillators, $W_i$ are the weights and $R$ is the molar gas constant. Above ~10 K the Eq. (5) describes well the experimental data with energies and weights of the Bose-Einstein oscillators listed in the inset of Fig. A6. The energies of the oscillators compare well with characteristic low-energy terahertz-infrared absorption bands associated with water in beryl. Especially the sharp absorption band at ~117 cm$^{-1}$ is well reproduced and accounts for the lower temperature contributions to the heat capacity of the crystal water. Below ~10 K, the approach with four Bose-Einstein oscillators is not fully adequate to describe the heat capacity completely. The difference $\delta C_p(T)/T$ between the calculated (on the basis of four oscillators) and the experimental data is plotted in the lower inset in Fig. A6. This difference is reminiscent of the heat capacity of a two-level *Schottky* anomaly with the two levels at an energy distance of ~10 K. Such an anomaly indicates excitations of very low energy. The weight, however, is very small and corresponds to ~2% of such two-level systems per formula unit of beryl also leaving extrinsic effects like lattice defects or imperfections which are modified by the annealing process as a possible origin.

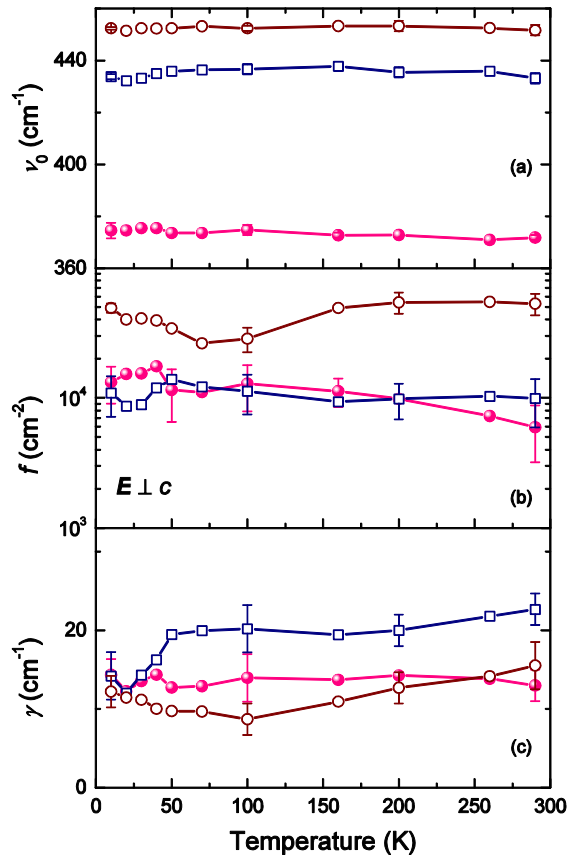

FIG. A3. (Color online). Detailed temperature dependences of parameters of the water-related absorption lines in beryl observed in the far-infrared range for polarizations $E\perp c$: eigenfrequencies $\nu_0$, oscillator strengths $f$, damping factors $\gamma$ and coupling constant.

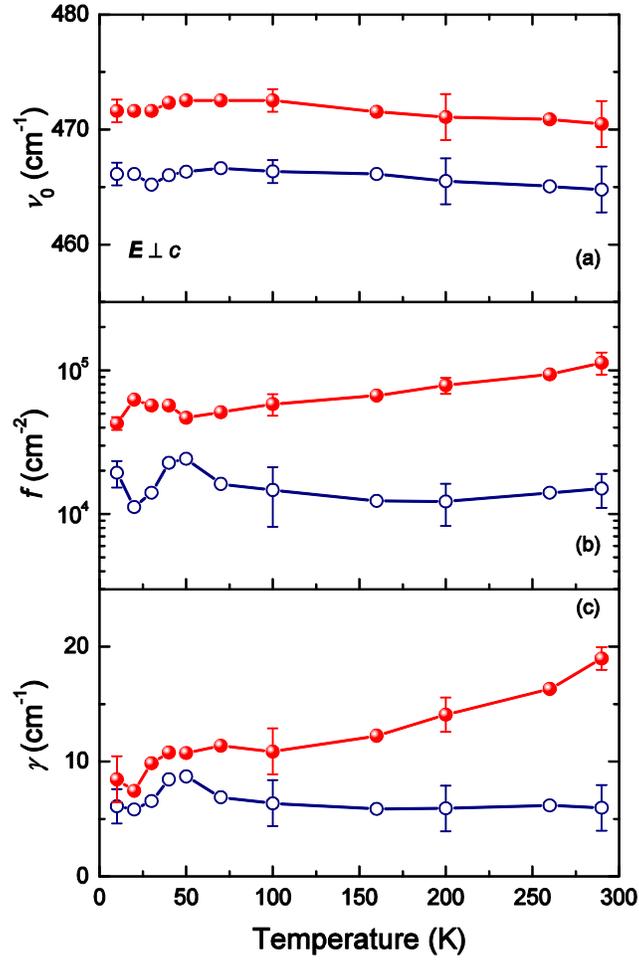

FIG. A4. (Color online). Detailed temperature dependences of parameters of the water-related absorption lines in beryl observed in the far-infrared range for polarizations $E\perp c$: eigenfrequencies $\nu_0$, oscillator strengths $f$, damping factors $\gamma$ and coupling constant.

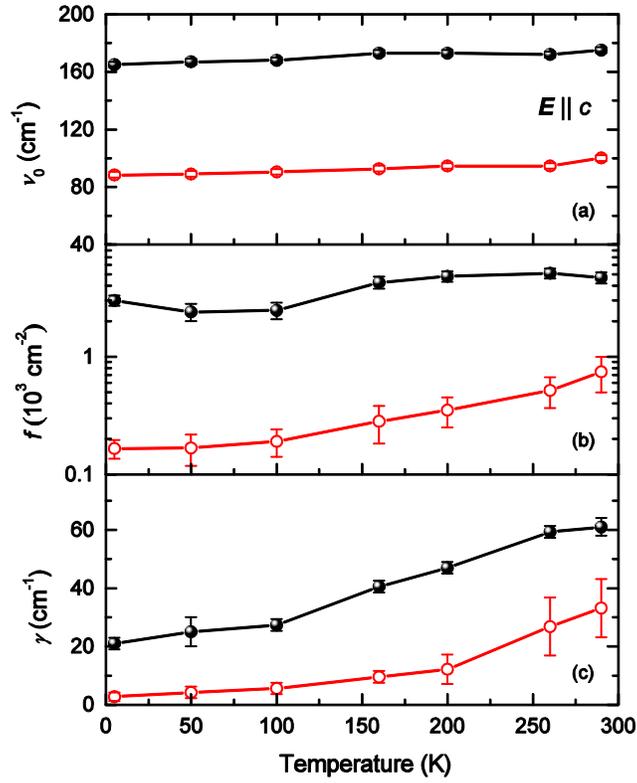

FIG. A5. (Color online). Detailed temperature dependences of parameters of the water-related absorption lines in beryl observed in the far-infrared range for polarizations E∥c: eigenfrequencies $\nu_0$, oscillator strengths $f$, damping factors $\gamma$ and coupling constant.

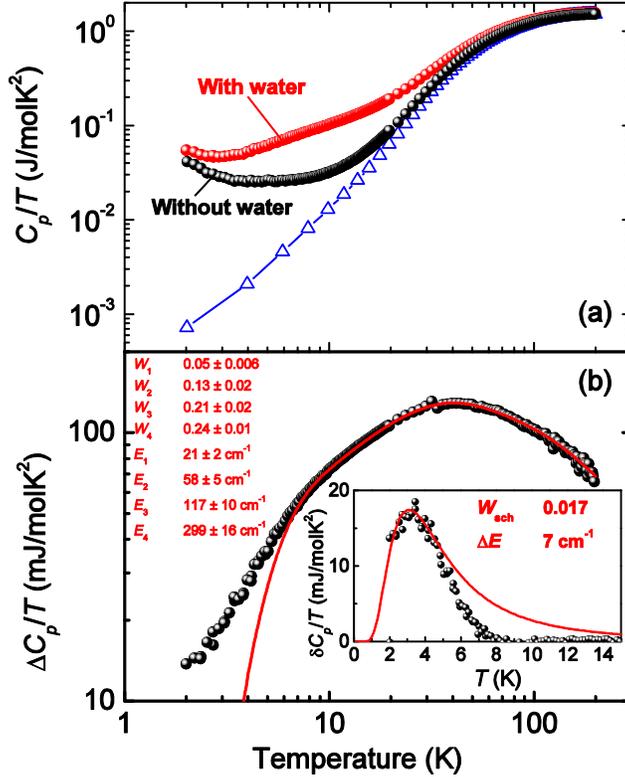

FIG. A6. (Color online). (a) Heat capacity of beryl with (red circles) and without (black circles) crystal water. Literature data by Hemingway et al. [81] for a beryl crystal with 0.36$H_2O$ per formula unit are presented by blue triangles. (b) Difference of the heat capacities, $\Delta C_p(T)/T$, of a beryl crystal with and without crystal water. The solid (red) line shows the fit of the data with the heat capacity of four Bose-Einstein oscillators (expression (A1)) with energies E (in wavenumbers) and weight factors W. The difference $\delta C_p(T)/T$ between the fit and the experimental data is shown by open symbols in the inset together with the heat capacity of a two-level Schottky anomaly (red line in the inset) with level distance of $\Delta E = 7$ cm$^{-1}$ and with the weight $W_{Sch} = 0.017$.

---

[1] L.D. Gelb, K.E. Gubbins, R. Radhakrishnan, and M. Sliwinska-Bartkowiak, Rep. Prog. Phys. **62**, 1573 (1999).

[2] H. Dosch, Appl. Surface Science **182**, 192 (2001).

[3] F. Franks, *Water, a Matrix of Life* (Royal Society of Chemistry, Cambridge, United Kingdom, 2000), 2nd ed.

[4] S. Solomon, *Water: The Epic Struggle for Wealth, Power, and Civilization* (Harper, New York, 2010).